\documentclass[12pt]{article}
\usepackage{bm,amsmath,amssymb,graphicx,cases}

\begin{document}
\pagenumbering{arabic}
\begin{titlepage}

\title{Cosmic structure formation in massive conformal gravity}

\author{F. F. Faria$\,^{1,*}$ and G. S. Silva$\,^{2,3,\dagger}$ \\ \\
$^1$\, Centro de Ci\^encias da Natureza, \\ 
Universidade Estadual do Piau\'i, \\
64002-150 Teresina, PI, Brazil. \\
$^2$\, Campus prof. Ant\^onio Geovanne 
Alves de Sousa, \\ Universidade Estadual do Piau\'i, \\
64260-000 Piripiri, PI, Brazil. \\
$^3$\, Campus Caxias, \\ Universidade Estadual do Maranh\~ao, \\ 
65604-380 Caxias, MA, Brazil.}

\date{}
\maketitle

\begin{abstract}
We study the evolution of cosmological density 
perturbations in massive conformal gravity (MCG). We show that conformal 
fermionic matter alone does not generate growing density perturbations 
in the subhorizon regime. By introducing a conformally coupled scalar 
field, however, an effective cold dark matter component emerges at the 
effective level. The resulting coupled perturbation equations admit growing 
solutions with a stronger effective gravitational interaction, allowing 
cosmic structure formation in MCG and potentially favoring earlier structure 
formation at high redshifts.
\end{abstract}

\thispagestyle{empty}
\vfill
\bigskip
\noindent * felfrafar@hotmail.com \par
\noindent $\dagger$ giofisica21@gmail.com \par
\end{titlepage}
\newpage


\section{Introduction}
\label{sec1}


The formation and evolution of large-scale structures in the universe 
\cite{Lif,Pee,Wei1} provide one of the most stringent tests of any 
cosmological model. While the standard $\Lambda$CDM model successfully explains 
a wide range of observations, including the cosmic microwave background (CMB) 
\cite{Hin,Pla}, baryon acoustic oscillations (BAO) \cite{Eis,Ala,Bos}, and 
large-scale structure (LSS) \cite{Hey,Des1}, some important problems 
suffered by it, such as the cosmological constant problem 
\cite{Wei2,Car}, the Hubble tension between the early and late universe observational 
data \cite{Ver,DiV,Rie}, and recent tensions in the growth rate of structures and 
the value of $\sigma_8$ \cite{Asg,Boy,Pou,Pan}, have motivated the exploration of 
alternative cosmological theories. We propose MCG, which is a conformally invariant 
theory of gravity in which the gravitational action is the sum of the Weyl action 
with the Einstein-Hilbert action conformally coupled to a scalar dilaton field 
\cite{Far4}, as a potential candidate to address such challenges.

The study of MCG is motivated by the search for a gravitational theory that 
simultaneously possesses consistent high-energy and low-energy limits. At the 
quantum level, the conformal coupling between the dilaton and the Ricci 
scalar ensure the unitarity of the theory  despite the presence of a massive 
spin-$2$ ghost field \cite{Far1}, while the Weyl-squared term provides the 
higher-derivative structure necessary for renormalizability \cite{Far2,Far3}. 
After spontaneous conformal symmetry breaking, the theory reproduces general 
relativity at low energies \cite{Far9}. In addition, it has been shown so far 
that the MCG cosmological model fits well with the 
Type Ia supernovae (SNIa) data without the cosmological constant problem 
\cite{Far5}, predicts the observed primordial abundances of light elements 
\cite{Far6}, passes the generalized second law of thermodynamics test 
\cite{Far7} and has a big bang singularity that is quantumly harmless 
\cite{Far3}. However, the problem of cosmic structure formation in MCG has 
not yet been fully explored.

In this work, we investigate the evolution of cosmological density 
perturbations in MCG and analyze whether the theory can support the growth 
of structures. We first derive the relativistic continuity and Euler equations 
for a conformal fermionic fluid directly from the conservation of the 
energy-momentum tensor. We then obtain the perturbed MCG cosmological equations
and derive the corresponding linear perturbation equation.

Our analysis shows that a universe filled only with conformal fermionic matter 
does not allow gravitational clustering in the subhorizon regime because the 
relativistic pressure term dominates the perturbation dynamics, leading to 
oscillatory solutions instead of growing modes. To address this problem, we 
introduce an additional conformally coupled matter scalar field $\phi$. Although 
the full theory remains fundamentally conformally invariant, the rapid oscillatory 
regime of $\phi$ generates an effective pressureless component 
that behaves as cold dark matter (CDM). At the effective level, this induces a
non-vanishing trace in the averaged energy-momentum tensor and dynamically 
shifts the spacetime away from the fundamental $R=0$ regime of the theory.

We show that the effective perturbation equations admit asymptotic growing 
solutions during the epoch dominated by the effective CDM component. 
In particular, the effective CDM density fluctuations possess a growing mode 
$\delta_\phi \propto (1+z)^{-1.886}$, which grows faster than the standard 
$\Lambda$CDM result $\delta_m \propto (1+z)^{-1}$. In addition, the conformal 
fermionic perturbations acquire a growing particular solution 
$\delta \propto (1+z)^{-0.886}$ induced by the gravitational coupling to the 
effective CDM sector. Consequently, the scalar field $\phi$ generates 
gravitational potential wells that subsequently attract conformal fermions, 
providing a natural mechanism for structure formation in MCG.

The resulting growth history differs qualitatively from that of a universe filled 
exclusively with conformal fermionic matter, where perturbations remain oscillatory 
at subhorizon scales. The enhanced growth found in the effective CDM era may favor 
earlier formation of gravitationally bound structures at high redshifts. Nevertheless, 
a quantitative comparison with observations requires the computation of cosmological 
observables such as the growth factor $D(z)$, the growth rate $f\sigma_8(z)$, halo 
abundances, and the matter power spectrum, which are left for future work.

This paper is organized as follows. In Sec. 2, we review the classical equations 
of MCG. In Sec. 3, we derive the relativistic hydrodynamic equations for the 
conformal perfect fluid. In Sec. 4, we obtain the perturbed MCG cosmological 
equations. In Sec. 5, we derive the perturbation equation for conformal fermionic
 matter and show the absence of growing modes. In Sec. 6, we introduce the 
conformally coupled matter scalar field and demonstrate how it generates an effective 
CDM component. In Sec. 7, we derive the corresponding linear growth 
equation for the effective CDM fluctuations and analyze its implications for 
structure formation. Finally, in Sec. 8, we present our conclusions.


\section{Classical MCG}
\label{sec2}


The total action of MCG is given by\footnote{Here we consider natural units 
in which $c=\hbar = 1$.} \cite{Far8}
\begin{equation}
S = \int{d^{4}x} \, \sqrt{-g}\bigg[ \varphi^{2}R 
+ 6 \partial^{\mu}\varphi\partial_{\mu}\varphi
- \frac{1}{2\alpha^2} C^2 \bigg] 
+ \int{d^{4}x\mathcal{L}_{m}},
\label{1}
\end{equation}
where $\varphi$ is a gravitational scalar field called 
dilaton, $\alpha$ is a dimensionless coupling 
constant,
\begin{equation}
C^2 = C^{\alpha\beta\mu\nu}C_{\alpha\beta\mu\nu} = R^{\alpha\beta\mu\nu}
R_{\alpha\beta\mu\nu} - 2R^{\mu\nu}R_{\mu\nu} + \frac{1}{3}R^{2}
\label{2}
\end{equation}
is the Weyl curvature invariant,
 $R^{\alpha}\,\!\!_{\mu\beta\nu} 
= \partial_{\beta}\Gamma^{\alpha}_{\mu\nu} + \cdots$ is the Riemann tensor, 
$R_{\mu\nu}$ is the Ricci tensor, $R$ is the scalar curvature, and 
$\mathcal{L}_{m} = \mathcal{L}_{m}[g_{\mu\nu},\Psi]$
is the conformally invariant Lagrangian density of the matter field 
$\Psi$.   

Varying the action (\ref{1}) with respect to $g^{\mu\nu}$ and $\varphi$, 
we find the MCG field equations\footnote{Since 
both $B_{\mu\nu}$ and $T_{\mu\nu}$ are traceless, we can see that the 
field equation (\ref{4}) follows from the trace of (\ref{3}) and thus 
contains no new dynamical information.}
\begin{equation}
\varphi^{2}G_{\mu\nu}
+  6 \partial_{\mu}\varphi\partial_{\nu}\varphi - 3g_{\mu\nu}\partial^{\rho}
\varphi\partial_{\rho}\varphi + g_{\mu\nu} \Box
\varphi^{2} - \nabla_{\mu}\nabla_{\nu}\varphi^{2} 
- \alpha^{-2}B_{\mu\nu} = \frac{1}{2} T_{\mu\nu},
\label{3}
\end{equation}
\begin{equation}
\left(\Box - \frac{1}{6}R\right)\varphi = 0,
\label{4}
\end{equation}
where
\begin{equation}
G_{\mu\nu} = R_{\mu\nu} - \frac{1}{2}g_{\mu\nu}R
\label{5}
\end{equation}
is the Einstein tensor,
\begin{equation}
B_{\mu\nu} = \left(\nabla^{\lambda}\nabla^{\rho} 
- \frac{1}{2}R^{\lambda\rho}\right)C_{\mu\lambda\nu\rho}
\label{6}
\end{equation}
is the Bach tensor,
\begin{equation}
T_{\mu\nu} = - \frac{2}{\sqrt{-g}} \frac{\delta \mathcal{L}_{m}}
{\delta g^{\mu\nu}}
\label{7}
\end{equation}
is the conformal matter energy-momentum tensor, and
$\Box = \nabla^{\mu}\nabla_{\mu}$.

At this point, it is important to remark that the symmetries underlying 
the theory, namely the general coordinate invariance and the conformal 
invariance, naturally allow the appearance of a quartic potential term 
$\lambda_\varphi\int{d^4x\sqrt{-g}\varphi^4}$ in the gravitational sector 
of the total MCG action (\ref{1}). In spite of this possibility, such a term 
is not taken into account in MCG because its inclusion implies that the 
Minkowski metric ceases to be a vacuum solution of the field equations, 
thus rendering the conventional $S$-matrix approach inconsistent. Although 
the potential term naturally arises in the loop divergences of the MCG 
effective action, we can cancel such divergences by considering the 
renormalized value of the dimensionless coupling constant $\lambda_\varphi$ 
equal zero.

Even though the gravitational sector of the theory doesn't have a 
classical potential, quantum corrections of the dilaton field generate an 
one-loop effective potential that has a minimum value $\varphi_0$ away
from the origin \cite{Mat}, which spontaneously breaks the conformal 
symmetry of the theory via the Coleman-Weinberg mechanism \cite{CW}. The 
consistency of the theory with solar system tests imposes that the 
spontaneously broken vacuum expectation value of the dilaton field 
be $\varphi_{0} = \sqrt{3/32\pi G} \approx 2\times 10^{18} \, \mbox{GeV}$ 
\cite{Far9}. In this case, we find 
that the MCG field equations (\ref{3}) and (\ref{4}) reduces to
\begin{equation}
G_{\mu\nu} - m^{-2} B_{\mu\nu} 
= \frac{16\pi G}{3} \, T_{\mu\nu},
\label{8}
\end{equation}
\begin{equation}
R = 0,
\label{9}
\end{equation}
at the classical fundamental level, where $m =\varphi_0\alpha$ is the 
effective mass of the massive spin-$2$ field (ghost) with negative energy 
that appears in the theory in addition to the usual massless spin-$2$ field 
(graviton) with positive energy.

In addition, for $\varphi = \varphi_0$, the conformally invariant MCG 
line element 
$ds^2 = \left(\varphi/\varphi_{0}\right)^{2}g_{\mu\nu}dx^{\mu}dx^{\nu}$  
reduces to the general relativistic line element  
\begin{equation}
ds^2 = g_{\mu\nu}dx^{\mu}dx^{\nu},
\label{10}
\end{equation}
and the conformally invariant MCG geodesic equation
\begin{equation}
\frac{d^{2}x^{\lambda}}{d\tau^2} + \Gamma^{\lambda}\,\!\!_{\mu\nu}
\frac{dx^{\mu}}{d\tau}\frac{dx^{\nu}}{d\tau} +\frac{1}{\varphi}
\frac{\partial\varphi}{\partial x^{\rho}} \left( g^{\lambda\rho} + 
\frac{dx^{\lambda}}{d\tau}\frac{dx^{\rho}}{d\tau}\right) = 0
\label{11}
\end{equation}
reduces to the general relativistic geodesic equation
\begin{equation}
\frac{d^{2}x^{\lambda}}{d\tau^2} + \Gamma^{\lambda}_{\mu\nu}
\frac{dx^{\mu}}{d\tau}\frac{dx^{\nu}}{d\tau} = 0,
\label{12}
\end{equation}
where
\begin{equation}
\Gamma^{\lambda}_{\mu\nu} = \frac{1}{2}g^{\lambda\rho}\left( 
\partial_{\mu}g_{\nu\rho} + \partial_{\nu}g_{\mu\rho} 
- \partial_{\rho}g_{\mu\nu} \right)
\label{13}
\end{equation}
is the Levi-Civita connection. The full classical fundamental content of MCG 
can be obtained from (\ref{8}), (\ref{9}), (\ref{10}) and (\ref{12}) without 
loss of generality. 

Before proceeding, it is worth to compare MCG with another conformally 
invariant theory of gravity called conformal gravity (CG), whose action is 
given by \cite{Man1,Man2}
\begin{equation}
S = - \frac{1}{2\alpha^2}\int{d^{4}x} \, \sqrt{-g} 
\left(C^{\alpha\beta\mu\nu}C_{\alpha\beta\mu\nu} \right)
+ \int{d^{4}x\mathcal{L}_{m}},
\label{14}
\end{equation}
where $\mathcal{L}_{m}$ is the same conformally invariant matter Lagrangian 
density from MCG. By varying (\ref{14}) with respect to $g_{\mu\nu}$, we 
obtain the field equation
\begin{equation}
B_{\mu\nu} = -\frac{\alpha^2}{2}T_{\mu\nu}.
\label{15}
\end{equation}
By comparing (\ref{15}) with (\ref{8}) and (\ref{9}), 
we can see that despite sharing local conformal invariance, MCG and CG represent 
distinct gravitational theories with different dynamics.


\section{Relativistic conformal hydrodynamic equations}
\label{sec3}


The variation of the conformally invariant matter Lagrangian density \cite{Man3}
\begin{equation}
\mathcal{L}_{m} = -\sqrt{-g}\bigg[S^{2}R + 6\partial^{\mu}S\partial_{\mu}S 
+ \lambda_S S^{4}  + \frac{i}{2}\left(\, \overline{\psi}
\gamma^{\mu}D_{\mu}\psi - D_{\mu}\overline{\psi}\gamma^{\mu}\psi \right) 
- \mu S\overline{\psi}\psi\bigg]
\label{16}
\end{equation}
with respect to $S$, $\overline{\psi}$ and $\psi$ gives the field equations
\begin{equation}
\left(12\Box - 2R \right) S 
- 4\lambda_S S^3 + \mu \overline{\psi}\psi = 0,
\label{17}
\end{equation}
\begin{equation}
i\gamma^{\mu}D_{\mu}\psi - \mu S \psi = 0,
\label{18}
\end{equation}
\begin{equation}
iD_{\mu}\overline{\psi}\gamma^{\mu} + \mu S \overline{\psi} = 0,
\label{19}
\end{equation}
where $S$ is a scalar Higgs field, $\lambda_S$ and $\mu$ are 
dimensionless coupling constants, $\psi$ is the fermion field,  
$\overline{\psi} = \psi^{\dagger}\gamma^{0}$ is the adjoint fermion field, 
$D_{\mu} = \partial_{\mu} 
+ [\gamma^{\nu},\partial_{\mu}\gamma_{\nu}]/8 - [\gamma^{\nu},\gamma_{\lambda}]
\Gamma^{\lambda}_{\mu\nu}/8$, and $\gamma^{\mu}$ 
are the general relativistic Dirac matrices, which satisfy the anticommutation 
relation $\{\gamma^{\mu},\gamma^{\nu}\} = 2g^{\mu\nu}$.

Substituting (\ref{16}) into (\ref{7}), and using (\ref{17})-(\ref{19}), 
we obtain the energy-momentum tensor
\begin{eqnarray}
T_{\mu\nu} &=& 8\partial_{\mu}S\partial_{\nu}S - 2g_{\mu\nu}\partial^{\rho}S
\partial_{\rho}S- 4 S\nabla_{\mu}\nabla_{\nu} S  
+ g_{\mu\nu}S\Box S \nonumber \\ &&
+ \, 2S^{2}\left(R_{\mu\nu} - \frac{1}{4}g_{\mu\nu}R\right) 
+ T^{f}_{\mu\nu},
\label{20}
\end{eqnarray}
where
\begin{equation}
T^{f}_{\mu\nu} = \frac{i}{4}\big(\, \overline{\psi}
\gamma_{\mu}D_{\nu}\psi - D_{\nu}\overline{\psi}\gamma_{\mu}\psi 
+ \overline{\psi}\gamma_{\nu}D_{\mu}\psi - D_{\mu}\overline{\psi}\gamma_{\nu}
\psi \big) - \frac{1}{4} g_{\mu\nu}\mu S\overline{\psi}\psi
\label{21}
\end{equation}
is the conformal fermion energy-momentum tensor.

Considering that the Higgs field 
acquires the spontaneously broken constant vacuum expectation value 
$S_{0} \approx 246 \, \mbox{GeV}$, and taking an incoherent average 
of $T^{f}_{\mu\nu}$ over all the fermionic modes propagating in a FLRW background, 
we find that (\ref{20}) becomes the energy-momentum tensor of the conformal perfect 
fluid that fills the MCG universe
\cite{Far6} 
\begin{equation}
T_{\mu\nu} = 2S_0^2\left(R_{\mu\nu} - \frac{1}{4}g_{\mu\nu}R\right)
+ \left( \rho + p \right)u_{\mu}u_{\nu} + g_{\mu\nu}p 
+ g_{\mu\nu}\rho_{\Lambda},
\label{22}
\end{equation}
where $\rho$ is the mass density of the conformal fluid, 
$p$ is the pressure of the conformal fluid, $u^{\mu}$ is the four-velocity 
of the fluid, which is normalized to $u^{\mu}u_{\mu} = - 1$, and 
$\rho_{\Lambda}$ is the vacuum energy (dark energy) density, with 
\begin{equation}
p = 0, \qquad \qquad \rho_{\Lambda} = \frac{1}{4}\rho,
\label{23}
\end{equation}
for a non-relativistic conformal fluid, and
\begin{equation}
p = \frac{1}{3}\rho, \qquad \qquad \rho_{\Lambda} = 0,
\label{24}
\end{equation}
for a relativistic conformal fluid.

By considering (\ref{23}) and (\ref{24}), we can write (\ref{22}) in the form 
\begin{equation}
T_{\mu\nu} = 2S_0^2\left(R_{\mu\nu} 
- \frac{1}{4}g_{\mu\nu}R\right) + \left(u_{\mu}u_{\nu} 
+ \frac{1}{4}g_{\mu\nu}\right)\left(1 + w \right)\rho,
\label{25}
\end{equation}
where $w = p/\rho$ is the equation-of-state parameter. It is not difficult to 
see that (\ref{25}) is traceless, which ensures that (\ref{9}) remain valid 
at the effective level. In addition, we can explicitly see from (\ref{25}) that 
the vacuum energy density with $w = -1$ does not contribute to the 
dynamic evolution of the MCG universe, which solves the cosmological constant 
problem found in the $\Lambda$CDM model.

Substituting the scalar-perturbed FLRW metric in the conformal newtonian gauge
\begin{equation}
ds^2 = a^{2}(\eta)\left[-\left(1+2\Phi\right)d\eta^2 
+ \left(1-2\Psi\right)\gamma_{ij}dx^idx^j\right],
\label{26}
\end{equation}
the perturbed four-velocity $u^{\mu} = a^{-1}(1 - \Phi,v^i)$ and the 
perturbed energy density $\rho(\eta,\textbf{r}) = \bar{\rho}(\eta) 
+ \delta{\rho}(\eta,\textbf{r})$ into the fermionic part of (\ref{25}), 
which is given by
\begin{equation}
T^{f}_{\mu\nu} =  \left(u_{\mu}u_{\nu} 
+ \frac{1}{4}g_{\mu\nu}\right)\left(1 + w \right)\rho,
\label{27}
\end{equation}
and keeping only the terms up to first order in the perturbations, 
we obtain\footnote{From now on, we drop the superscript $f$, for simplicity.}
\begin{eqnarray}
T^{0}\,\!\!_{0} &=& -\,\frac{3}{4}(1+w)(\bar{\rho} + \delta\rho), 
\label{28} \\
T^{0}\,\!\!_{i} &=& \gamma_{ij}v^{j}(1+w)\bar{\rho}, 
\label{29} \\
T^{i}\,\!\!_{0} &=& - \, v^{i}(1+w)\bar{\rho}, 
\label{30} \\
T^{i}\,\!\!_{j} &=& \frac{1}{4}\delta^{i}\,\!\!_{j}(1+w)(\bar{\rho}
+\delta\rho), 
\label{31}
\end{eqnarray}
where $a(\eta)$ is the conformal scale factor, $d\eta = dt/a$ is the 
conformal time, $\Phi = \Phi(\eta,\textbf{r})$ and 
$\Psi = \Psi(\eta,\textbf{r})$ are scalar metric perturbations, $v^{i} 
= dx^{i}/d\eta$ is the conformal velocity perturbation, $\delta\rho$ is the 
energy density perturbation, and $\gamma_{ij}$ is the spatial metric of 
constant curvature $K$.

The substitution of the metric (\ref{26}) into (\ref{13}) gives the 
up to first-order terms
\begin{eqnarray}
\Gamma^{0}_{00} &=& \mathcal{H} + \Phi', 
\label{32} \\
\Gamma^{0}_{0i} &=& \partial_i \Phi, 
\label{33} \\
\Gamma^{0}_{ij} &=& \gamma_{ij}\left[ \mathcal{H} - \Psi' - 2\mathcal{H}
\left(\Phi+\Psi\right)\right], 
\label{34} \\
\Gamma^{i}_{00} &=& \gamma^{ij}\partial_j \Phi, 
\label{35} \\
\Gamma^{i}_{j0} &=& \delta^{i}\,\!\!_{j}\left(\mathcal{H} 
- \Psi'\right), 
\label{36} \\
\Gamma^{i}_{jk} &=& \Gamma^{i}_{jk}(\gamma) 
- \delta^{i}\,\!\!_{j}\partial_k \Psi - \delta^{i}\,\!\!_{k}\partial_j \Psi 
+ \gamma_{jk}\gamma^{il}\partial_l \Psi, 
\label{37}
\end{eqnarray}
where $\mathcal{H} = a'/a$ is the conformal Hubble parameter and the prime 
denotes $\partial/\partial\eta$. Using (\ref{28})-(\ref{37}) in the conservation 
law
\begin{equation}
\nabla_\mu T^{\mu}\,\!\!_{0} = \partial_\mu T^{\mu}\,\!\!_{0} 
+ \Gamma^{\mu}_{\mu\lambda}T^{\lambda}\,\!\!_{0} 
- \Gamma^{\lambda}_{\mu 0}T^{\mu}\,\!\!_{\lambda} = 0,
\label{38}
\end{equation}
and separating the zeroth-order and first-order terms, we obtain the 
conformal continuity equation
\begin{equation}
\bar{\rho}\,' + 4\mathcal{H}\bar{\rho} = 0,
\label{39}
\end{equation}
and the conformal evolution equation of the density perturbation
\begin{equation}
\delta\rho' + \frac{4}{3}\bar{\rho}\,\nabla \cdot \textbf{v} 
+ 4\mathcal{H}\delta\rho - 4\bar{\rho}\,\Psi' = 0,
\label{40}
\end{equation}
where $\nabla \cdot \textbf{v} = \partial_i v^i + \Gamma^{i}_{ij}(\gamma) v^j$.
Finally, the use of (\ref{39}) in (\ref{40}) gives the relativistic version of 
the continuity equation for a conformal perfect fluid
\begin{equation}
\delta' + \frac{4}{3}\left(\nabla \cdot \textbf{v} - 3\Psi'\right) = 0,
\label{41}
\end{equation}
where
\begin{equation}
\delta = \frac{\delta\rho}{\bar{\rho}}
\label{42}
\end{equation}
is the fractional density perturbation.

Now, making the substitutions of (\ref{28})-(\ref{37}) into the conservation law
\begin{equation}
\nabla_\mu T^{\mu}\,\!\!_{i} = \partial_\mu T^{\mu}\,\!\!_{i} 
+ \Gamma^{\mu}_{\mu\lambda}T^{\lambda}\,\!\!_{i} 
- \Gamma^{\lambda}_{\mu i}T^{\mu}\,\!\!_{\lambda} = 0,
\label{43}
\end{equation}
and neglecting second-order terms, we find
\begin{equation}
\bar{\rho}\,'\gamma_{ij}v^j + \bar{\rho}\,\gamma_{ij}v'^{j}
+ \frac{1}{4}\partial_i (\delta\rho) + \bar{\rho}\, \partial_i \Phi 
+ 4\mathcal{H}\bar{\rho}\,\gamma_{ij}v^{j}  = 0.
\label{44}
\end{equation}
The additional use of (\ref{39}) then gives
\begin{equation}
\textbf{v}' = - \frac{1}{4}\nabla\delta - \nabla\Phi,
\label{45}
\end{equation}
which is the relativistic version of the Euler equation for a conformal 
perfect fluid.

Taking the conformal time derivative $\partial/\partial\eta$ of (\ref{41}) 
and the divergence of (\ref{45}), we find
\begin{equation}
\delta'' + \frac{4}{3}\left(\nabla \cdot \textbf{v}' - 3\Psi''\right) = 0,
\label{46}
\end{equation}
\begin{equation}
\nabla \cdot \textbf{v}' = - \frac{1}{4}\nabla^2\delta - \nabla^2\Phi.
\label{47}
\end{equation}
Finally, the combination of (\ref{46}) with (\ref{47})  gives
\begin{equation}
\delta'' - \frac{1}{3} \nabla^2\delta - \frac{4}{3}\nabla^2\Phi - 4\Psi'' = 0,
\label{48}
\end{equation}
which is the relativistic perturbation equation for a conformal perfect fluid. 
In addition to (\ref{48}), we also need to find the field equations of the MCG 
gravitational perturbations in order to describe the dynamics of the 
matter fluctuations in the theory.


\section{Perturbed MCG cosmological equations}
\label{sec4}


By substituting (\ref{26}) and (\ref{32})-(\ref{37}) into the Ricci tensor
\begin{equation}
R_{\mu\nu} = \partial_{\lambda}\Gamma^{\lambda}_{\mu\nu} 
- \partial_{\nu}\Gamma^{\lambda}_{\mu\lambda} 
+ \Gamma^{\lambda}_{\lambda\rho}\Gamma^{\rho}_{\mu\nu}
- \Gamma^{\rho}_{\mu\lambda}\Gamma^{\lambda}_{\nu\rho},
\label{49}
\end{equation}
the scalar curvature $R = g^{\mu\nu}R_{\mu\nu}$ and the Bach tensor (\ref{6}), 
and keeping only up to first-order terms, we obtain
\begin{eqnarray}
R_{00} &=& -\, 3\mathcal{H}' + \nabla^2\Phi + 3\mathcal{H}(\Phi'+\Psi') + 3\Psi'', 
\label{50} \\
R_{0i} &=& 2\nabla_i\left(\Psi' + \mathcal{H}\Phi\right),  
\label{51} \\
R_{ij} &=& \gamma_{ij} \big[ \mathcal{H}' + 2\mathcal{H}^2 + 2K - \Psi'' 
+ \nabla^2\Psi - 2(\mathcal{H}' + 2\mathcal{H}^2 + 2K)(\Phi + \Psi) 
\nonumber \\ &&   - \, \mathcal{H}\Phi' -5\mathcal{H}\Psi' \big] 
+ \nabla_i \nabla_j (\Psi - \Phi),
\label{52} \\
a^2R &=& - \, 6(\mathcal{H}'+\mathcal{H}^2+K) + 2\nabla^2\Phi - 4\nabla^2\Psi 
+12(\mathcal{H}'+\mathcal{H}^2+K)\Phi \nonumber \\ &&
+ \, 6\Psi'' + 6\mathcal{H}(\Phi' + 3\Psi'), 
\label{53} \\
a^2 B_{00} &=& \frac{2}{3}\left(\nabla^{2} + 3K \right)\nabla^2(\Phi + \Psi), 
\label{54} \\
a^2 B_{0i} &=& \frac{2}{3}\nabla_i\left(\nabla^2 + 3K\right)\left(\Phi' 
+ \Psi'\right), 
\label{55} \\
a^2 B_{ij} &=& \frac{1}{3}\left[ \gamma_{ij}\nabla^2 \left(\nabla^2 +2K 
- \partial_{\eta}^2\right) -\nabla_i \nabla_j \left(\nabla^2 - 3\partial_{\eta}^2
\right)\right](\Phi + \Psi), 
\label{56}
\end{eqnarray}
where both the covariant derivative $\nabla_i$ and the Laplacian $\nabla^2$ are 
associated with the spatial metric $\gamma_{ij}$.

Using (\ref{28})-(\ref{31}) and (\ref{50})-(\ref{56}) in (\ref{8}) and 
(\ref{9}), we obtain the zeroth-order MCG cosmological equations\footnote{Since 
$S_0^2 \ll 3/32\pi G$, we can neglect the first term of (\ref{25}) 
and use only its fermionic part (\ref{27}).}
\begin{equation}
\mathcal{H}' = - \frac{4\pi G}{3} a^2(1+w)\bar{\rho},
\label{57}
\end{equation}
\begin{equation}
\mathcal{H}' + 2\mathcal{H}^2 + 2K = \frac{4\pi G}{3} a^2(1+w)\bar{\rho},
\label{58}
\end{equation}
\begin{equation}
\mathcal{H}'+\mathcal{H}^2 + K = 0,
\label{59}
\end{equation}
and the first-order MCG cosmological equations
\begin{equation}
2\nabla^2\Psi - 6\mathcal{H}\Psi'
- \frac{2}{3m^2a^2}\left(\nabla^{2} + 3K \right)\nabla^2(\Phi + \Psi)
= 4\pi G a^2(1+w)\bar{\rho}(\delta + 2\,\Phi),
\label{60}
\end{equation}
\begin{equation}
\nabla_i\left[ \left(\Psi' + \mathcal{H}\Phi\right) 
- \frac{1}{3m^2a^2}\left(\nabla^2 + 3K\right)\left(\Phi' 
+ \Psi'\right)\right] 
= -\frac{8\pi G}{3} a^2(1+w)(\bar{\rho}\,\gamma_{ij}v^j),
\label{61}
\end{equation}
\begin{equation}
2\nabla^2\Phi - 4\nabla^2\Psi + 6\Psi'' + 6\mathcal{H}(\Phi' + 3\Psi') = 0,
\label{62}
\end{equation}
where we also used (\ref{59}) to find (\ref{57}), (\ref{58}) and (\ref{62}).

Subtracting (\ref{57}) from (\ref{58}), we obtain
\begin{equation}
\mathcal{H}^2 + K = \frac{4\pi G}{3} a^2(1+w)\bar{\rho},
\label{63}
\end{equation}
which determines the dynamics of the MCG universe. Although (\ref{63}) is
consistent with SNIa data for an open ($K = -1$) MCG universe \cite{Far5}, it 
still needs to be confronted with other cosmological data, which is beyond the 
scope of this paper. 


\section{Linear growth of conformal matter fluctuations}
\label{sec5}


In order to describe the formation of cosmic structures in MCG, we start by 
substituting the plane wave solutions
\begin{equation}
\delta(\eta,\textbf{r}) = \delta_k(\eta)e^{i\textbf{k}\cdot\textbf{r}}, 
\qquad
\Phi(\eta,\textbf{r}) = \Phi_k(\eta)e^{i\textbf{k}\cdot\textbf{r}},
\qquad
\Psi(\eta,\textbf{r}) = \Psi_k(\eta)e^{i\textbf{k}\cdot\textbf{r}}
\label{64}
\end{equation}
into (\ref{48}), (\ref{60}) and (\ref{62}), which gives
\begin{equation}
\delta'' + \frac{1}{3} k^2\delta +\frac{4}{3}k^2\Phi -4\Psi''= 0,
\label{65}
\end{equation}
\begin{equation}
-2k^2\Psi-6\mathcal{H}\Psi'-\frac{2k^2(k^2-3K)}{3m^2a^2}(\Phi+\Psi)
= 4\pi Ga^2(1+w)\bar{\rho}(\delta+2\Phi),
\label{66}
\end{equation}
\begin{equation}
-2k^2\Phi+4k^2\Psi +6\Psi''+6\mathcal{H}(\Phi'+3\Psi')=0,
\label{67}
\end{equation}
where $k = |\textbf{k}|$ is the wavenumber of the wave.

By imposing the quasi-static subhorizon limit
\begin{equation}
k^2\gg \mathcal{H}^2, \qquad k^2 \gg |K|,\qquad
\Phi',\Psi',\Phi'', \Psi'' = 0,
\label{68}
\end{equation}
in (\ref{65})-(\ref{67}), we find\footnote{Unlike general 
relativity, where scalar perturbations satisfy $\Phi = \Psi$ in 
the absence of anisotropic stress, MCG predicts the modified relation (\ref{71}). 
This difference originates from $R=0$ 
and may lead to observational signatures in gravitational lensing and large-scale 
structure.}
\begin{equation}
\delta'' + \frac{1}{3} k^2\delta +\frac{4}{3}k^2\Phi = 0,
\label{69}
\end{equation}
\begin{equation}
-2k^2\Psi-\frac{2k^4}{3m^2a^2}(\Phi+\Psi)
= 4\pi Ga^2(1+w)\bar{\rho}\,\delta,
\label{70}
\end{equation}
\begin{equation}
\Phi = 2\Psi,
\label{71}
\end{equation}
where we neglected the term proportional to $\Phi$ on the right-hand side of 
(\ref{66}) to find (\ref{70}) because density fluctuations dominate over the 
potential in the subhorizon. 

The substitution of (\ref{71}) into (\ref{70}) 
gives the modified Poisson equation
\begin{equation}
k^2\left(1+\frac{k^2}{m^2a^2}\right)\Phi =
-4\pi Ga^2(1+w)\bar{\rho}\,\delta.
\label{72}
\end{equation}
Finally, using (\ref{72}) and $\delta'' = a^2\ddot{\delta} 
+ a\dot{a}\dot{\delta}$ in (\ref{69}), we obtain
\begin{equation}
\ddot{\delta} + H\dot{\delta} + \left[\frac{k^2}{3a^2} 
-4\pi G_\mathrm{eff}(a,k)(1+w)\bar{\rho}\right] \delta = 0,
\label{73}
\end{equation}
where $H = \dot{a}/a$ is the Hubble parameter and
\begin{equation}
G_\mathrm{eff}(a,k) = \frac{4G}{3\left(1+\frac{k^2}{m^2a^2}\right)}
\label{74}
\end{equation}
is an effective gravitational constant. 

Considering that $m \gtrsim 10^{-2} \, \mbox{eV}$, according to solar system
tests \cite{Far9} and Cavendish like experiments \cite{Cav}, we can see that
\begin{equation}
\frac{k^2}{m^2a^2} \ll 1
\label{75}
\end{equation}
in the subhorizon limit, so that (\ref{73}) reduces to
\begin{equation}
\ddot{\delta} + H\dot{\delta} + \left[\frac{k^2}{3a^2} 
-\frac{16\pi G}{3}(1+w)\bar{\rho}\right] \delta = 0
\label{76}
\end{equation}
on the scales where the cosmic structures are formed. 
Using (\ref{63}), we can write (\ref{76}) in the form
\begin{equation}
\ddot{\delta} + H\dot{\delta} + \left[\frac{k^2}{3a^2} 
- 4\left(\frac{\mathcal{H}^2}{a^2} + \frac{K}{a^2}\right)\right] \delta = 0.
\label{77}
\end{equation}
Therefore, in the subhorizon limit, the pressure term $k^2/3a^2$ dominates 
and (\ref{77}) reduces to
\begin{equation}
\ddot{\delta} + H\dot{\delta} + \frac{k^2}{3a^2}\delta = 0,
\label{78}
\end{equation}
which is valid for both relativistic and non-relativistic conformal 
fermionic matter due to the independence that it has from $w$. 

The solution to (\ref{78}) oscillates inside the 
sound horizon rather than growing monotonically, which means that cosmic 
structures will never be formed in MCG if we consider that the matter 
content of the MCG universe is composed only by conformal fermions. 
Fortunately, this scenario changes if we consider the presence of 
conformal bosonic matter in the theory, as we will show next.

\section{Effective cold dark matter}
\label{sec6}

In order to address the problem of cosmic structure formation in MCG, 
let us introduce an extra matter scalar field $\phi$, which does not acquire 
a vacuum expected value. By conformally coupling this scalar field 
only with the metric and the Higgs field, we can extend the matter Lagrangian 
density (\ref{16}) by the conformally invariant term
\begin{equation}
\mathcal{L}_{\phi} = -\sqrt{-g}\left[\phi^2 R + 6 \partial^\mu \phi 
\partial_\mu \phi + \lambda_\phi \phi^4 + 2\lambda_{S\phi} S^2 
\phi^2 \right],
\label{79}
\end{equation}
where $\lambda_\phi$ and $\lambda_{S\phi}$ are dimensionless coupling 
constants.

The variation of (\ref{79}) with respect to $\phi$ gives the field equation
\begin{equation}
\left(6\Box - R \right) \phi 
- 2\lambda_\phi \phi^3 - 2\lambda_{S\phi}S^2\phi = 0.
\label{80}
\end{equation}
In addition, substituting (\ref{79}) into (\ref{7}), we obtain the conformal 
energy-momentum tensor
\begin{eqnarray}
T_{\mu\nu}^{\phi} &=& 12\partial_{\mu}\phi\partial_{\nu}\phi
- 6g_{\mu\nu}\partial^{\rho}\phi\partial_{\rho}\phi + 2 \left( g_{\mu\nu} 
\Box - \nabla_{\mu}\nabla_{\nu}+ G_{\mu\nu} \right)\phi^2 \nonumber \\ &&
- \, g_{\mu\nu}\left(\lambda_\phi \phi^4 + 2\lambda_{S\phi} S^2 
\phi^2 \right).
\label{81}
\end{eqnarray}
The additional use of (\ref{80}) shows that (\ref{81}) is traceless on-shell.

Considering that the Higgs field $S$ acquires the spontaneously broken constant 
vacuum expectation value $S_{0}$, and that 
$\phi$ is a homogeneous field in the FRLW background, we find that (\ref{80}) 
and (\ref{81}) become
\begin{equation}
\ddot{\phi} + 3H\dot{\phi} + \left(\dot{H} +2H^2 + \frac{K}{a^2} 
+ m_\phi^2 \right)\phi + \frac{1}{3}\lambda_\phi \phi^3   = 0,
\label{82}
\end{equation}
\begin{equation}
T^\phi_{00} = 6\dot\phi^2 +12H\phi\dot\phi 
+ 6\phi^2\left(H^2+\frac{K}{a^2}\right) +\lambda_\phi\phi^4
+6m_\phi^2\phi^2
\label{83}
\end{equation}
\begin{equation}
T^\phi_{ij} = \left[-2\phi^2\left(2\dot H+3H^2+\frac{K}{a^2}\right)
+2\dot\phi^2 -4\phi\ddot\phi -8H\phi\dot\phi -\lambda_\phi\phi^4
-6m_\phi^2\phi^2\right]a^2\gamma_{ij},
\label{84}
\end{equation}
where $m_\phi^2 = \lambda_{S\phi}S_0^2/3$ is the effective 
mass of $\phi$. 

The use of (\ref{83}) and (\ref{84}) allows us to write  $T^\phi_{\mu\nu}$ 
in the perfect fluid form
\begin{equation}
T^{\phi}_{\mu\nu} = \left( \rho_\phi + p_\phi \right)
u_{\mu}u_{\nu} + g_{\mu\nu}p_\phi,
\label{85}
\end{equation}
where $u^\mu = (1,0,0,0)$ is the comoving four-velocity,
\begin{equation}
\rho_\phi = 6\dot\phi^2 +12H\phi\dot\phi 
+ 6\phi^2\left(H^2+\frac{K}{a^2}\right) +\lambda_\phi\phi^4
+6m_\phi^2\phi^2
\label{86}
\end{equation}
is the density of the fluid, and
\begin{equation}
p_\phi = -2\phi^2\left(2\dot H+3H^2+\frac{K}{a^2}\right)
+2\dot\phi^2 -4\phi\ddot\phi -8H\phi\dot\phi -\lambda_\phi\phi^4
-6m_\phi^2\phi^2
\label{87}
\end{equation}
is the pressure of the fluid.

In the fast oscillation regime
\begin{equation}
m_\phi^2 \gg R,  \ \ \ \ \ \ \ m_\phi^2 \gg \lambda_\phi \phi^2,
\label{88}
\end{equation}
the field equation (\ref{82}) reduces to
\begin{equation}
\ddot{\phi} + 3H\dot{\phi} + m_\phi^2 \phi = 0,
\label{89}
\end{equation}
whose WKB solution is given by
\begin{equation}
\phi(t) = A a^{-3/2}\cos(m_\phi t),
\label{90}
\end{equation}
where $A$ is a normalization constant. 

Taking the time average of (\ref{90}), we obtain
\begin{equation}
\langle \dot\phi^2 \rangle = m_\phi^2 \langle \phi^2 \rangle, \qquad
\langle \phi \dot\phi \rangle = -\frac{3}{2}H \langle \phi^2 \rangle.
\label{91}
\end{equation}
Finally, using (\ref{88}) and (\ref{91}) in (\ref{86}) and (\ref{87}), we find
\begin{equation}
\rho_\phi = 12 m_\phi^2 \langle \phi^2 \rangle \propto a^{-3}, 
\qquad  p_\phi = 0,
\label{92}
\end{equation}
which means that $\phi$ behaves as CDM at the effective 
level\footnote{Rapidly oscillating scalar fields are known to behave 
effectively as pressureless matter after averaging over oscillation periods 
much shorter than the Hubble time \cite{Tur,Pre1,Amen,Mar}.}. It follows from 
(\ref{92}) that we can write (\ref{85}) in the reduced form 
\begin{equation}
T^\phi_{\mu\nu} = \rho_\phi u_\mu u_\nu.
\label{93}
\end{equation}
It is not difficult to see that the trace of (\ref{93}) is given by
\begin{equation}
T^\phi = g^{\mu\nu} T^\phi_{\mu\nu} =  - \rho_\phi,
\label{94}
\end{equation}
which implies that the introduction of $\phi$ breaks the conformal symmetry 
of the theory at the effective level. As a consequence, (\ref{93}) 
backreacts on the spacetime, shifting the geometry from the fundamental 
$R=0$ regime to an effective $R \neq 0$ regime.

By perturbing (\ref{93}), substituting the result into (\ref{38}) and 
(\ref{43}), and making some algebra, we obtain the effective CDM continuity 
equation
\begin{equation}
\dot{\bar{\rho}}_\phi + 3H\bar{\rho}_\phi = 0,
\label{95}
\end{equation}
and the effective CDM relativistic perturbation equation
\begin{equation}
\ddot{\delta}_\phi + 2H\dot{\delta}_\phi - \frac{1}{a^2}\nabla^2\Phi 
- 3\ddot{\Psi} - 6H\dot{\Psi} = 0.
\label{96}
\end{equation}
Since the presence of (\ref{93}) leads to $R \neq 0$, the MCG cosmological 
equations (\ref{57})-(\ref{62}) are no longer valid if we include the effective 
CDM in the theory, which changes both the dynamics of the MCG universe and those
of the scalar metric perturbations $\Phi$ and $\Psi$ that we must use in 
(\ref{96}). These new dynamics are explored in the next section.


\section{Linear growth of effective CDM fluctuations}
\label{sec7}


By considering both (\ref{27}) and (\ref{93}) in (\ref{8}),  we find
\begin{equation}
G_{\mu\nu} - m^{-2} B_{\mu\nu} = \frac{16\pi G}{3}\left[\left(u_{\mu}u_{\nu} 
+ \frac{1}{4}g_{\mu\nu}\right)\left(1 + w \right)\rho 
+ \rho_\phi u_\mu u_\nu\right].
\label{97}
\end{equation}
The perturbations of (\ref{97}) in the FRLW background 
give the zeroth-order modified MCG cosmological equation
\begin{equation}
H^2 + \frac{K}{a^2} = \frac{16\pi G}{9}\left[\frac{3}{4}
(1+w)\bar{\rho} + \bar{\rho}_\phi \right],
\label{98}
\end{equation}
and the first-order modified MCG cosmological equations
\begin{equation}
2\nabla^2\Psi - 6H\dot{\Psi} - \frac{2}{3m^2 a^2}
(\nabla^2 + 3K)\nabla^2(\Phi + \Psi) = \frac{16\pi G a^2}{3}\left[\frac{3}{4}
(1+w)\bar{\rho}\,\delta + \bar{\rho}_\phi \delta_\phi \right] ,
\label{99}
\end{equation}
\begin{equation}
\frac{2}{a^{2}}\nabla^{2}\Phi - \frac{4}{a^{2}}\nabla^{2}\Psi
+ 12\left(\dot H + 2H^{2} + \frac{K}{a^{2}}\right)\Phi + 6\ddot{\Psi}
+ 24H\dot{\Psi} + 6H\dot{\Phi} 
= \frac{16\pi G}{3}\bar{\rho}_{\phi}\delta_{\phi},
\label{100}
\end{equation}
where we neglected the term proportional to $\Phi$ on the right-hand side of 
(\ref{99}).

Substituting the plane waves solutions of $\delta$, $\delta_\phi$, $\Phi$ and $\Psi$ 
into (\ref{96}), (\ref{99}) and (\ref{100}), and imposing the subhorizon limits
(\ref{68}) and (\ref{75}), we find
\begin{equation}
\ddot{\delta}_\phi + 2H\dot{\delta}_\phi + \frac{k^2}{a^2}\Phi = 0,
\label{101}
\end{equation}
\begin{equation}
2k^2\Psi
= -\frac{16\pi G a^2}{3}\left[\frac{3}{4}
(1+w)\bar{\rho}\,\delta + \bar{\rho}_\phi \delta_\phi \right],
\label{102}
\end{equation}
\begin{equation}
k^2\Phi = 2k^2\Psi - \frac{8\pi G}{3} a^2 \bar{\rho}_\phi \delta_\phi.
\label{103}
\end{equation}
By combining (\ref{102}) with (\ref{103}), we obtain the Poisson equation
\begin{equation}
k^2\Phi = - 4\pi G a^2\left[(1+w)\bar{\rho}\,\delta  + 2\bar{\rho}_\phi \delta_\phi\right].
\label{104}
\end{equation}
The insertion of (\ref{104}) into (\ref{69}) and (\ref{101}) then gives
\begin{equation}
\ddot{\delta} + H\dot{\delta} + \frac{k^2}{3a^2}\delta
= \frac{16\pi G}{3}\left[(1+w)\bar{\rho}\,\delta + 2\bar{\rho}_\phi \delta_\phi \right],
\label{105}
\end{equation}
\begin{equation}
\ddot{\delta}_\phi + 2H\dot{\delta}_\phi  = 
 4\pi G\left[(1+w)\bar{\rho}\,\delta + 2\bar{\rho}_\phi \delta_\phi \right],
\label{106}
\end{equation}
which determine the coupled evolution of linearized conformal fermions and 
effective CDM density perturbations in the MCG universe.

Although (\ref{105}) and (\ref{106}) do not have general exact solutions, 
we can find approximated asymptotic solutions for each era of the MCG universe. 
Using the solutions to 
the continuity equations (\ref{39}) and (\ref{95}), which are given by
\begin{equation}
\bar{\rho} = \bar{\rho}_0\left(\frac{a_0}{a}\right)^4, 
\qquad \bar{\rho}_\phi = \bar{\rho}_{\phi 0}\left(\frac{a_0}{a}\right)^3,
\label{107}
\end{equation}
and considering that the MCG universe is open ($K = -1$) \cite{Far5}, we can 
write (\ref{98}) in the form
\begin{equation}
\dot a^2 = \frac{A}{a^2} + \frac{B}{a} + 1,
\label{108}
\end{equation}
where
\begin{equation}
A = \frac{4\pi G}{3} (1+w)\rho_0 a_0^4, \qquad 
B = \frac{16\pi G}{9}\rho_{\phi 0}a_0^3,
\label{109}
\end{equation}
with the subscript $0$ denoting values at the present time $t_0$.

The solution to (\ref{108}) interpolates between
\begin{equation}
a = \left(2\sqrt{A}\,t\right)^{1/2}
\label{110}
\end{equation}
at early times, when the radiation term $A/a^2$ dominates,
\begin{equation}
a =\left(\frac{3}{2}\sqrt{B}\,t\right)^{2/3}
\label{111}
\end{equation}
at intermediate times, when the effective CDM term $B/a$ dominates, and
\begin{equation}
a = t
\label{112}
\end{equation}
at late times, when the curvature term dominates.

Using (\ref{107}), (\ref{110})-(\ref{112}) and $a = 1/(1+z)$ in 
(\ref{105}) and (\ref{106}), we find the approximated asymptotic solutions
\begin{equation}
\delta = A\cos{\left(\frac{C_1 k}{1+z}\right)} 
+ B\sin{\left(\frac{C_1k}{1+z}\right)}, \qquad 
\delta_\phi = C_2 - C_3\ln{\left(1+z\right)},
\label{113}
\end{equation}
in the radiation era,
\begin{equation}
\delta = \delta_h + C_1(1+z)^{-0.886} + C_2(1+z)^{3.386}, \qquad 
\delta_\phi =  C_3(1+z)^{-1.886} + C_4(1+z)^{2.386},
\label{114}
\end{equation}
in the effective CDM era, and
\begin{equation}
\delta = A\cos{\left[\frac{k}{\sqrt{3}}\ln{(1+z)}\right]} 
- B\sin{\left[\frac{k}{\sqrt{3}}\ln{(1+z)}\right]}, \qquad 
\delta_\phi = C_1 + C_2\left(1+z\right),
\label{115}
\end{equation}
in the curvature era, where
\begin{equation}
\delta_h = (1+z)^{1/4}\left[A\cos{\left(\frac{C_5k}{\sqrt{1+z}}\right)} 
+ B\sin{\left(\frac{C_5k}{\sqrt{1+z}}\right)} \right],
\label{116}
\end{equation}
and $C_i$ ($i = 1,...,5$) are integration constants.

The asymptotic solutions obtained above provide a clear physical picture of
structure formation in MCG. In the radiation era, we can see from 
(\ref{113}) that the conformal fermionic perturbations oscillate as acoustic 
waves, while the effective CDM perturbations grow only logarithmically as
$\delta_\phi \propto -\ln(1+z)$. This behavior is analogous to the 
M\'esz\'aros effect \cite{Mes} in the standard cosmological scenario, 
where dark matter perturbations experience only slow growth during the 
radiation-dominated epoch. Consequently, no significant nonlinear structures 
are formed at this stage.

A qualitatively different regime emerges during the effective CDM era. In this
case, the solutions (\ref{114}) show that the effective CDM perturbations possess 
a growing mode $\delta_\phi \propto (1+z)^{-1.886}$, which grows substantially 
faster than the standard $\Lambda$CDM result $\delta_m \propto (1+z)^{-1}$ 
\cite{Dod}. As a consequence, effective CDM overdensities may reach the nonlinear 
regime at earlier times. The conformal fermionic perturbations also acquire a 
growing particular solution $\delta \propto (1+z)^{-0.886}$ driven by the effective 
CDM fluctuations in addition to the oscillatory homogeneous mode $\delta_h$. Although 
this growth is slower than that of the effective CDM component, it implies that the 
conformal fermion distribution is gravitationally dragged by the effective CDM halos. 
Therefore, the scalar CDM field forms the potential wells first, while the 
conformal fermions subsequently accumulate inside them.

Finally, in the curvature-dominated era, it follows from (\ref{115}) that the 
growth of perturbations effectively ceases. The effective CDM fluctuations approach 
a constant value  while the conformal fermionic perturbations remain oscillatory. 
At this stage, the previously formed halos are already present, and the role of 
gravity is mainly to preserve the existing structures rather than to generate 
new growth.

Therefore, the mechanism of structure formation in MCG can be summarized as
follows. During the radiation era, density fluctuations remain small 
and grow only logarithmically. During the effective CDM era, the scalar CDM field 
behaves as pressureless matter and develops rapidly growing density perturbations 
that generate halos. These halos subsequently capture conformal fermions through 
gravitational attraction. Once curvature domination begins,
the growth of perturbations freezes out and the previously formed structures
survive. The enhanced growth rate predicted by (\ref{114}) may therefore favor
earlier galaxy formation than in the standard $\Lambda$CDM model \cite{Pre2,Whi,Spr}, 
potentially alleviating some tensions associated with high-redshift galaxies
observed by the James Webb Space Telescope (JWST) \cite{Men,Ada,Lab}.

\section{Final remarks}
\label{sec8}

In this work, we have investigated the formation of cosmic structures within
the framework of MCG. Starting from the conservation
of the energy-momentum tensor, we derived the relativistic continuity and Euler
equations for a conformal fermionic fluid and obtained the corresponding linear
perturbation equation.

We showed that, if the matter content of the universe is described solely by
a conformal fermionic fluid, density perturbations exhibit oscillatory behavior
in the subhorizon regime, preventing the formation of cosmic structures. This
result is a direct consequence of the dominance of the relativistic pressure
term proportional to $k^2/3a^2$. To overcome this limitation, we introduced 
an extra conformally invariant scalar field $\phi$. After spontaneous 
symmetry breaking, the scalar field acquires an effective mass and enters a 
rapidly oscillating regime. Using a covariant WKB treatment, we demonstrated 
that its averaged energy-momentum tensor behaves as a pressureless fluid with 
energy density $\rho_\phi \propto a^{-3}$, thus providing an effective CDM 
component. Although the fundamental theory remains 
conformally invariant, the effective energy-momentum tensor acquires a 
non-vanishing trace, dynamically shifting the spacetime from the fundamental 
$R=0$ sector to an effective $R \neq 0$ regime. This modification changes both 
the background cosmological equations and the perturbation dynamics.

The resulting coupled perturbation equations admit asymptotic analytical 
solutions in the three different eras of the MCG universe. During the 
radiation era, the effective CDM perturbations grow only logarithmically, 
reproducing a behavior analogous to  the M\'esz\'aros effect. During the effective 
CDM era, however, the effective CDM  perturbations 
develop a rapidly growing mode $\delta_\phi \propto (1+z)^{-1.886}$, while the 
conformal fermionic perturbations acquire a growing particular solution 
$\delta \propto (1+z)^{-0.886}$, in addition to an oscillatory homogeneous 
component. These results indicate that the scalar CDM field forms gravitational 
potential wells which subsequently attract conformal fermions, thereby providing 
a viable mechanism for the formation of galaxies and large-scale structures in MCG. 
At late times, when the curvature term dominates the cosmic expansion, the growth 
of perturbations effectively freezes. The effective CDM fluctuations approach a constant 
value and the conformal fermionic perturbations remain oscillatory. Therefore, the 
structures generated during the effective CDM era survive into the late universe.

The enhanced growth rate predicted by MCG may favor earlier structure formation 
than in the standard $\Lambda$CDM model. However, the viability of this scenario 
cannot be established solely from the asymptotic solutions derived here. A 
complete assessment requires solving the coupled cosmological system numerically 
and confronting the model with observables such as the growth factor $D(z)$, the 
growth rate $f\sigma_8(z)$, weak-lensing measurements, galaxy clustering data, 
and the matter power spectrum, which can be directly compared with measurements 
from Planck, eBOSS, DES and Euclid \cite{Pla,Bos,Des1,Euc}. Such an analysis 
will be presented elsewhere. Future work may also investigate nonlinear structure 
formation, spherical collapse, halo mass functions, vector and tensor perturbations, 
and possible observational signatures associated with the modified relation 
$\Phi = 2\Psi$ predicted by MCG.


\end{document}